\documentstyle[aps,epsf]{revtex}

\newcommand{\bee}{\begin{equation}}
\newcommand{\ee}{\end{equation}}
\newcommand{\bea}{\begin{eqnarray}}
\newcommand{\eea}{\end{eqnarray}}
\newcommand{\R}{\rm I\kern-.2emR}
\newcommand{\C}{\rm \kern.25em\vrule height1.4ex
depth-.12ex width.06em\kern-.31em C}
\newcommand{\N}{{\rm I\kern-.16em N}}
\newcommand{\Z}{{\rm Z\kern-.35em Z}}

\begin{document}
\draft
\twocolumn[\hsize\textwidth\columnwidth\hsize\csname
@twocolumnfalse\endcsname
\title{Quasi-asymptotic freedom in the two dimensional $O(3)$ model
}

\vskip 1.0truecm                                                                
\author{Adrian Patrascioiu}
\address{\it Physics Department, University of Arizona, \\
 Tucson, AZ 85721, U.S.A.}
\date{\today}
\maketitle
\begin{abstract}                                                                
The behavior of the renormalized spin 2-point function in the $O(3)$ 
and dodecahedron spin model are investigated numerically. The Monte
Carlo data show excellent agreement between the two models. The
short distance behavior comes very close to standard theoretical
expectations, yet it differs significantly from it. A possible explanation
of this situation is offered.
\end{abstract}                                                                  
\pacs{11.25.Bt, 11.15.Ha, 75.10.Hk}
]                                                                               
%111111111111111111111111111111111111111111111111111111111111111111111          
\narrowtext
\vskip2mm

Are two dimensional (2D) nonabelian $\sigma$-models asymptotically 
free? After more than two decades, the answer remains unclear, with
lots of supportive evidence on both sides of the issue. The main point
of this letter is to provide some fresh evidence that even though
the orthodox picture provides an excellent description of physics
at intermediate momemta and distances not too short, it is in fact
incorrect, a situation which I belive occurs also in $QCD_4$. 

To that end I will present the result of Monte Carlo (MC) investigations of
 the continuum limit of the spin 2-point function in four spin models:
 i)   Ising model,
 ii)  $O(2)$ model,
 iii) $O(3)$ model and
 iv)  dodecahedron spin model.
 As I will show, there is strong numerical evidence that the
continuum limit of the dodecahedron spin model is identical to that
of the $O(3)$ model. Undergoing a freezing transition at nonzero 
temperature, the latter is not likely to be asymptotically free. 
Moreover the continuum limit of the spin 2-point function of the
$O(3)$ model disagrees
with the PT formula with the value of $\Lambda/m$
predicted by Hasenfratz, Maggiore and Niedermayer
 based on the thermodynamic Bethe ansatz \cite{hmn} and the
normalization constant fixed by the scaling hypothesis of
 the form factor approach proposed by
 Balog and Niedermaier (BN) \cite{bn}. 

The numerical data suggest that in all the four models investigated,
the short distance behaviour of the spin 2-point function is of the
form $1/x^{1/4}$.

As a biproduct of my investigation, I report also on the behavior
of the critical exponent $\eta$, which probably equals 1/4 in all four models.
 The subject of
logarithmic corrections to $\eta=1/4$ in the $O(2)$ model has been
discussed in several recent papers \cite{ki},\cite{ps},
\cite{campo},\cite{janke}, since a theoretical prediction
stemming from the perturbative renormalization group approach 
exists \cite{kost} and it is not clear that the data are consistent
with it. As I will discuss shortly, the existence of a continuum limit
obtainable from the lattice model via multiplicative renormalization, an
 assumption which seems to be corroborated by the numerics, suggests that
no such logarithmic corrections are present.

I would like to begin by reviewing the intimate connection between
the short distance behavior of the continuum spin 2-point function and
the critical exponent $\eta$. First some definitions:
let $P=(p,0)$ and \\
- Spin 2-point function $G(p)$:
\bee
        G(p)={1\over L^2}\langle |\hat s(P)|^2\rangle;\ \
\hat s(P)=\sum_x e^{iPx} s(x)
\ee
- Magnetic susceptibility $\chi$:
\bee
       \chi=G(0)
\ee
- Correlation length $\xi$:
\bee
  \xi={1\over 2\sin(\pi/L)}\sqrt{(G(0)/G(1)-1)}
\ee
As the continuum spin 2-point function I will use the limit
$\xi\to\infty$ of
\bee
%       G(x/\xi)=\langle s(0).s(x)\rangle\xi^2/\chi\ \
       {G(x/\xi)\equiv {\langle s(0).s(x)\rangle\xi^2/\chi}}\ \
\ee
%Let $z$ be the physical continuum distance ($z\equiv $x/\xi$$, 
Let $z$ be the physical continuum distance ($z\equiv x/\xi$, 
limit $\xi\to\infty$);
the exponents $\bar \eta$ and $\bar r$ are defined by requiring 
that in the limit
$z\to0$ the expression
\bee
           G(z)z^{\bar \eta}{(log(z))}^{2\bar r}
\ee
is finite and different from 0. Let $\eta$ and $r$ be defined so that in
the limit $\xi\to\infty$ the quantity
\bee
           {c_{\it kt}\equiv {\chi\over {\xi^{2-\eta}{log(\xi)}^{-2r}}}}
\ee
is finite and different from 0.

I would like to claim that the existence of a continuum limit with the
short distance behaviour described by (5) requires $\eta=\bar \eta$ and
$r=-\bar r$. Indeed in the continuum limit, the limit $z\to0$ is  reached by
 letting both $x\to\infty$ and $\xi\to\infty$ but $x/\xi\to0$. In fact
the same consideration shows that in the same limit
the lattice spin 2-point function must have the property
\bee      
          \langle s(0).s(x)\rangle x^{\bar \eta}
\ee
goes to a finite nonzero constant (i.e. there are no logarithmic modifications).
 
The Kosterlitz \cite{kost} prediction is that $\bar r=-1/16$. Finite size
scaling at the {\it kt}-point do indicate a negative $r$ though smaller
in magnitude \cite{janke}. The paradox discussed by many \cite{ki},
\cite{campo},\cite{janke} is that the thermodynamic data suggest a
positive value for $r$ \cite{ps},\cite{janke}. As explained above this is 
consistent with the fact that $r=-\bar r$, rather than $r=\bar r$ as
incorrectly claimed in the literature.

However the fact that eq.(7) requires the lattice spin 2-point function
to decay as a pure power with no logarithmic corrections  
suggests that most likely $r=-\bar r=0$. Indeed while for $\xi<\infty$
eq.(7) is valid only for $x\ll\xi$, baring some nonuniformity, eq.(7)
ought to be describing the large distance behaviour of the lattice spin
2-point function at the critical point.  

Next I would like to discuss the main point of this paper, which concerns
the asymptotic behaviour of the continuum spin 2-point function
$G(z)$ at short distances. It is known  that in the Ising model $\eta=1/4$
 $r=0$. I stated above what perturbative renormalization group (RG) 
arguments predict for $O(2)$. For $O(3)$ similar arguments \cite{brezin}
predict $\eta=0$ $r=-1$. However when combined with the thermodynamic
Bethe ansatz and the scaling hypothesis in the
form factor approach, this latter prediction becomes 
even more precise
\cite{bn} :
\bea
G(z)&=&{1.000034657\over 3\pi^3\,1.001687^2}(t+ln(t)\cr
    &+&1.1159+ln(t)/t+.1159/t)^2,
\eea
  where  $t\equiv {-ln(z\Lambda)}, \Lambda=e/8.$\ \
I am not aware of any predictions regarding $\eta$ and $r$ for the
dodecahedron spin model.

\begin{figure}[htb]
\centerline{\epsfxsize=8.0cm\epsfbox{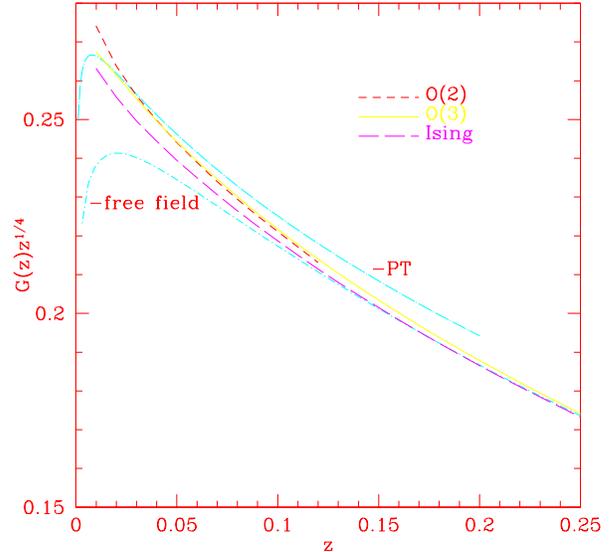}}
\caption{Spin 2-point function $G(z)$ versus physical distance $z$
 for the Ising, $O(3)$ and $O(2)$
models. Also shown the free field curve and the BN refined PT curve  
for $O(3)$.}
\label{fig:comp}
\end{figure}

In Fig.\ref{fig:comp} I present my Monte Carlo results for $G(z)z^{1/4}$ for
(in rising order) Ising, $O(3)$ and $O(2)$; I also show the curve 
for free field (lowest curve) and the PT curve with the normalization
constant predicted by Balog and Niedermaier  
 (upper curve) \cite{bn}.
The MC data were taken at $\xi \sim 167$ and L=1230. They represent
the on axis correlation and I will discuss shortly the lattice artefacts.
For now I would like to make the following observations:
\newline 1. \  \ In all three models $G(z)$ approaches the free field
value for $z\to\infty$, as required by the Orstein-Zernke behaviour
\cite{bf}.
\newline 2. \ \ For $z \geq .05$ $G(z)$ in the three models behaves
quite similarly.
\newline 3. \ \ The data suggest that in all three models $\eta=1/4$,
but that compared to the Ising model the short distance behavior may be
more singular in $O(2)$ and less singular in $O(3)$. This is 
would be consistent
with a negative $\bar r$ in $O(2)$ but a positive one in $O(3)$ 
(although the data are also consistent with $\bar r=0$).
\newline 4. \ \ Even though for $.02<z<.1$ 
the difference with the BN refined PT curve is small (about 2$\%$), it is 
clearly there and, as I will indicate next, it is not a lattice artefact.

\begin{figure}[htb]
\centerline{\epsfxsize=8.0cm\epsfbox{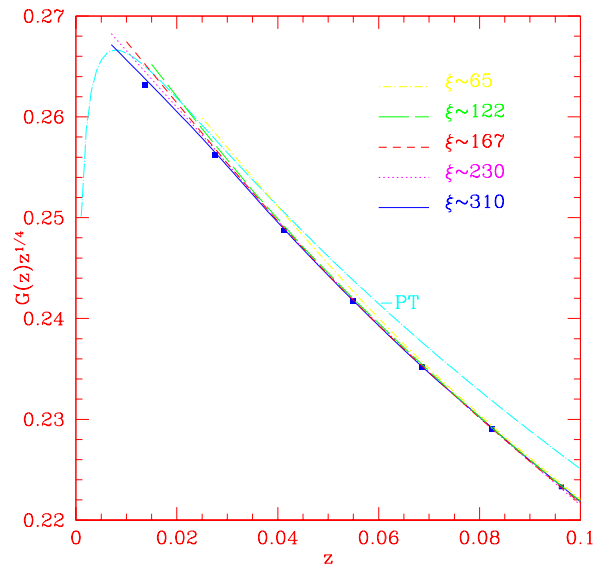}}
\caption{Spin 2-point function $G(z)$ versus physical distance $z$
 for the $O(3)$ 
model at several correlation lengths. 
Also shown the BN refined PT curve.}
\label{fig:th}
\end{figure}

In Fig.\ref{fig:th} I present the MC results for the $O(3)$ model for
 $\xi$ ranging from 65 to 310. The error
bars are about .3$\%$, too small to show. The data were produced using Wolff's
method \cite{w}.  At each $\beta$ and $L$ value, at least five independent runs
were made and the error estimated using the jack-knife method. The primary
source of error are the values of $\chi$ and $\xi$, whose determination involves
large distances. To give a feeling for the size of the lattice artefacts, 
for $\xi\sim 310$ the (solid) curve represents $G(z)$ on axis while the points
$G(z)$ along the diagonal. Both this test and the good agreement of these
data with the data at $\xi\sim 230$ suggest that, barring some very slow
drift, the continuum limit has been
reached for $z \geq .03$. 
The continuum values are clearly off the BN refined PT
curve (upper curve).

\begin{figure}[htb]
\centerline{\epsfxsize=8.0cm\epsfbox{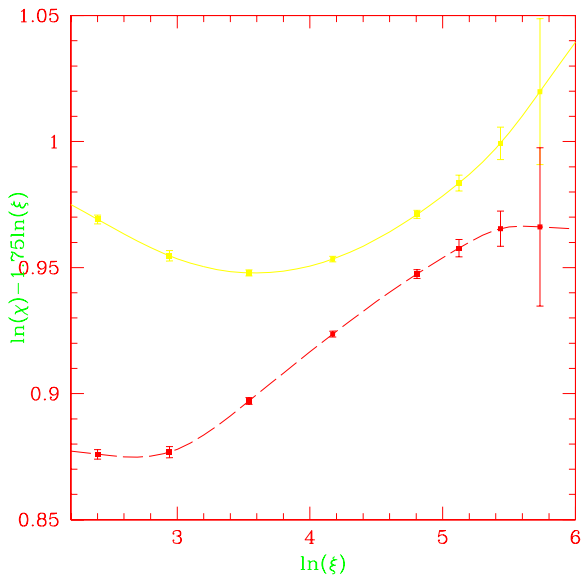}}
\caption{$ln(\chi)-1.75ln(\xi)$ versus $ln(\xi)$ for the $O(3)$
model. Dashed curve represents MC data combined with the BN bootstrap 
prediction,
a correct prediction requiring convergence to 1 for $\xi\to\infty$.}
\label{fig:chixi}
\end{figure}

As explained above, the critical behaviour of $\chi$ and $\xi$ carry
relevant information regarding the short distance behaviour of $G(z)$.
In Fig.\ref{fig:chixi}
 I represent $ln(\chi)-1.75ln(\xi)$. The upper curve represents
the MC data. The lower (dashed) curve represents the ratio of the
 values of this  
 quantity computed via the BN formula 
\bee
      {\chi\over \xi^2}={3\pi\,1.001687^2\over {4\,1.000034657\, \beta^2}} 
   \left\{1+{.1816\over \beta}+{.133\over \beta^2}+{.1362\over \beta^3}\right\}
\ee 
over its MC values.
A correct prediction of the BN formula would require the dashed curve to
approach 1 for $\xi\to\infty$. 
The data produce a (dashed) curve which is not flat and 
which seems to approache the line at 1 with a nonzero slope. 

The upper (solid) curve indicates that
$r$ could be different from 0 and in fact negative. To determine whether
there is a logarithmic modification to $\eta=1/4$ and obtain a 
precise value for $r$ one would need data at larger $\xi$. However
the lack of smoothness of the curve suggests that the errors are
underestimated already at $\xi\sim 310$ (probably due to critical slowing
down).

\begin{figure}[htb]
\centerline{\epsfxsize=8.0cm\epsfbox{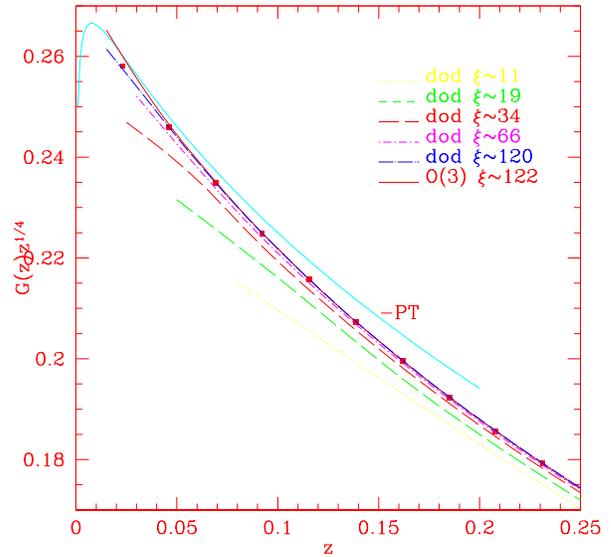}}
\caption{Spin 2-point function $G(z)$ in the dodecahedron model for
several correlation lengths. Also shown $G(z)$ for the $O(3)$ model
at $\xi\sim 122$ and the BN refined PT curve.}
\label{fig:dod}
\end{figure}

The last piece of numerical evidence which suggests that $O(3)$ is not
asymptotically free comes from a comparison of its spin 2-point 
function with that of of the dodecahedron spin model. This is shown in Fig.
\ref{fig:dod},
which shows $G(z)$ for the latter model at $\xi\sim 11,19,34,66$ and 120 
(broken line) and $G(z)$ for the $O(3)$ model at $\xi\sim 122$ 
(solid line on axis, dots on the diagonal). As the
correlation length increases, $G(z)$ for the dodecahedron model approaches 
that of $O(3)$ and at $\xi\sim 120$ the two curves are practically
indistinguishable for $z \geq .04$. It appears rather certain that
 the two models share the same $G(z)$ in the
continuum limit. However, as already mentioned, it is highly unlikely
that the dodecahedron model, undergoing freezing at finite $\beta$,
exhibits asymptotic freedom.

$\it Discussion$

1. I believe the numerical evidence presented above suggests that in
the $O(3)$ nonlinear $\sigma$ model the short distance behaviour of
$G(z)$ corresponds to $\bar \eta=1/4$, $\bar r \geq 0$. 

2. Perturbative renormalization group arguments predict that
in all $O(N)$ models $N \geq 3$ $\eta=0$, $\bar r$ is negative
and varies monotonically with $N$ from -1 for $O(3)$ to -1/2 for
$O(\infty)$. Based on the observed behavior of $O(3)$ I would
conjecture that for all $3 \leq N<\infty$ $\eta=1/4$.  
 Since the
spherical model $O(\infty)$ has $\eta=0$, $\bar r=-1/2$, such
a scenario would correspond to a nonuniformity of the $1/N$ expansion.
The uniformity of the $1/N$ expansion for $\beta \to \infty$ has been
questioned before \cite{1/n}.

3. Since for $.02\leq z \leq .1$ the deviation from the BN refined PT 
prediction 
is 2$\%$ or less, one may wonder why that is so?  I think the answer is
in Fig.1, where I compare $G(z)$ for different models. Please note
that the free field curve describes also the spherical model and
$G(z)$ for it is quite close to that of $O(3)$ for $z$ not too small. It is
well known \cite{kupi} that the $1/N$ expansion is legitimate at
fixed $\tilde \beta \equiv {\beta \over N}$ and the only issue
\cite{1/n} is whether this expansion is uniform for 
$\tilde \beta\to\infty$.
The reason the orthodoxy comes so close to the truth is two fold:
i) the fast
convergence of the $1/N$ expansion for moderate $\tilde \beta$
and ii) the large value of $\beta_{crt}$. Thus, even though I believe
both the $1/N$ expansion and perturbation theory are nonuniform, they
come very close to the truth
 at moderate correlation length. Since for distances not too small
or momenta not too large the continuum limit is well described at 
moderate $\xi$ values, the orthodoxy will describe correctly physics at
such distances and/or momenta. It is only at short distances or large
momenta that the true test of the orthodoxy can be performed.

4. Until now I discussed only some new numerical information. Some readers,
such as those who recently wrote {\it `We believe however that the
continuing accumulation of unambiguous, consistent and increasingly
accurate numerical support for the RG predictions from a variety of
independent approaches leaves liitle if any space for alternative pictures.'}
\cite{but},
may not find my numerical evidence conclusive. I will invite them then
to reconsider the two theoretical arguments advanced by Patrascioiu and Seiler 
regarding the existence of a massless phase in {\it all} $O(N)$ models:

i) We showed rigorously \cite{jsp} that if a certain cluster, baptized
the {\it {equatorial cluster}}, does not percolate, the model must be
massless. We also advanced some nonrigorous arguments why that
cluster did not seem likely to percolate \cite{pat}.
After seven years since those arguments were advanced,
no mathematical physicist has proved the contrary or given us an 
example of how this equatorial cluster might percolate (see 
{\it Open Problems in Mathematical Physics}, http://www.iamp.org). 
Can any skeptical
reader give such an example?

ii) We also showed rigorously that among smooth configurations, the 
dominant ones must be the {\it superinstantons} \cite{super} not the
much publicized instantons. But in a superinstanton configuration there is
a ring of arbitrary size in which the spin points roughly 
in a certain direction.
Hence in such a configuration, the inverse image of a small patch of the 
sphere will form a ring and hence the equatorial cluster will not percolate.
Therefore the existence of superinstantons as the dominant configuration
at weak coupling reinforces the argument for the existence of a massless
phase and hence the absence of asymptotic freedom in the massive phase.

I think irrespective of any numerics, to paraphrase Butera and Comi 
\cite{but}, `the two arguments mentioned above 
leave little if any space for the standard picture to be correct'.
  
5. Even though strictly incorrect, the standard picture provides an excellent
phenomenological description at distances not too small ($z>.02$) or
momenta not too large ($p<2\pi/.02\sim 300$. This is not something 
never seen before. Indeed it is taught in every course in Quantum Mechanics
that the probability to find a particle trapped in a potential well (say
an $\alpha$-particle) decays exponentially in time. The time constant of
that exponential is called the life time of the state, and innumerable
experiments over the years have measured such life times. I am not aware
of anybody ever having claimed to have detected deviations from this
alleged exponential law. Yet it is a mathematical fact that  if the particle
decays at all, the large time asymptotic behavior cannot be exponential, 
but some inverse power of
the time \cite{com}. This fact is little known in the physics community 
(witness the many papers by particle physicists computing the lifetime
of the `false cosmological vacuum')
because in fact the exponential law describes very well the data. My claim
is that a similar situation occurs in the 2D nonlinear $\sigma$ models
(and probably $QCD_4$): 
even though eventually, at $\beta$ sufficiently large superinstantons
win and render the model massless, at moderate values of $\beta$ such as 
$\beta=2$ for $O(3)$, instantons (localized defects) dominate the
typical configuration, which behaves very much as predicted by 
perturbation theory and/or the $1/N$ expansion. Only by probing sufficiently 
small distances or large momenta should one be able to detect major
deviations from the expected behavior. 
 
These ideas stem from my long term collaboration with Erhard Seiler and
were crystalized by my recent collaboration with J.Balog, M.Niedermaier,
F.Niedermayer and P.Weisz.

\end{document}